\documentclass[
 reprint,
nofootinbib,
 amsmath,amssymb,
 aps,prab,
floatfix,
]{revtex4-2}

\usepackage{graphicx}
\usepackage{dcolumn}
\usepackage{bm}
\usepackage[hyperfootnotes=false]{hyperref}
\usepackage{scrextend}
\usepackage{xcolor}

\usepackage{epstopdf} 
\AppendGraphicsExtensions{.tif}

\usepackage{booktabs}
\usepackage{subfigure}
\usepackage{bibunits}
\usepackage{multirow}
\usepackage{array}
\usepackage{float}

\usepackage{sidecap}

\usepackage{amsmath}
\usepackage{graphicx}
\usepackage{siunitx}

\begin{document}

\preprint{APS/123-QED}
\title{Plasma treated metals after H- irradiation and its effect on vacuum breakdown behaviour}

\author{ C. Serafim\textsuperscript{a,b}}\email{catarina.serafim@cern.ch}
\author{S. Calatroni\textsuperscript{a}}\email{sergio.calatroni@cern.ch}
\author{F. Djurabekova\textsuperscript{b}}
\author{M. C. Giordano\textsuperscript{a}}
\author{M. Himmerlich\textsuperscript{a}}
\author{V. Bjelland\textsuperscript{a,c}}
\author{C. Kouzios\textsuperscript{a}}
\author{P. Costa Pinto\textsuperscript{a}}
\author{A. T. Perez-Fontenla\textsuperscript{a}}
\author{W. Wuensch\textsuperscript{a}}
\author{A. Grudiev\textsuperscript{a}}
\author{S. Sgobba\textsuperscript{a}}

\affiliation{\textsuperscript{a}CERN, European Organization for Nuclear Research, 1211 Geneva, Switzerland;}
\affiliation{\textsuperscript{b}Helsinki Institute of Physics and Department of Physics, P.O. Box 43, FI-00014 University of Helsinki, Finland;}
\affiliation{\textsuperscript{c}Department of Physics, NTNU, 9491 Trondheim, Norway;}


\begin{abstract}
Vacuum breakdown in accelerator structures is a critical challenge that occurs under high electric fields. In environments subjected to hydrogen ion irradiation or high beam losses, such as in Radio-Frequency Quadrupoles (RFQ), residual hydrocarbons from the vacuum may result in carbon contamination of the metal surfaces from charged particle induced cracking. Under these conditions, it has been assessed that surface carbon contamination leads to a decrement of the surface electric field holding properties. This study extends the latest research by exploring the efficacy of Oxygen Plasma Cleaning (OPC) on metal electrodes irradiated by low energy hydrogen ion beam with the purpose of reducing surface carbon contamination. OPC treatment has been employed on different metals, namely copper beryllium (CuBe2), oxygen-free copper (Cu-OFE), and stainless steel (SS316LN). Treated electrodes have been tested for electric field performance in a DC pulsed system and results compared with non-irradiated electrodes and irradiated ones without OPC treatment. The study indicates a significant reduction in carbon contamination by OPC, enough to allow irradiated materials to achieve performances comparable with the electric field strength of raw surfaces. Moreover, it has been observed that stainless steel samples had some alteration in the surface chemistry that enhanced the material’s ability to sustain high electric fields while decreasing vacuum arcing events. Notably, OPC treated SS316LN electrodes surpassed the performance value of untreated ones, demonstrating the potential of plasma treatments in extending the operational performance of accelerator components.

\end{abstract}

\maketitle

\vspace{1cm}

\section{Introduction}

Vacuum arcing has been identified as a main concern in high gradient RF accelerators structures. The breakdown appearance is due to high electric fields between metal surfaces, as is the case for example of the CERN Radio-Frequency Quadrupole (RFQ),  the first element of the injection chain of the Large Hadron Collider (LHC). 
Multiple studies have tried to explain the appearance of vacuum breakdowns using both atomistic simulations and experimental testing \cite{Sergio_Taborelli_materials_RF, Descoeudres_Sergio_Taborelli, PhysRevSTAB_Taborelli, PhysRevAccelBeams_Anders_Flyura_Laura, PhysRevAccelBeams_Yinon_Inna, Classification_BD_anton, helga_kai_flyura_modelling_BD, roni_2024}.
According to the literature, breakdown appearance is dependent on material properties \cite{materials_chart_flyura, PhysRevSTAB.12.032001} making the choice of materials for RF applications a major factor to be exploited.
This work follows on the experiments conducted on dedicated test electrodes in regard to materials studies for breakdown resistance of irradiated surfaces. The main aim of those experiments, reported in \cite{paper1_catarina, paper2-arxiv-catarina}, was to find a more suitable material for RFQ manufacturing, since copper - material currently in use - has suffered some degradation over the years due to breakdowns and hydrogen implantation in the form of blistering \cite{lopez_serafim}, caused by the continuous operational exposure to a hydrogen ion beam.
While vacuum arcs are still yet to be fully comprehended, especially in terms of event's initiation, the current studies will hopefully bring new additional contributions to the field of research. 
Carbon contamination has been recognized to be the cause of decrease of electric field holding performance in irradiated electrodes \cite{paper2-arxiv-catarina}. It has been discovered that the surface becomes contaminated by a carbon layer, possibly due to the cracking of hydrocarbons present in the accelerator vacuum.  Carbon contamination has been a well-known issue during ion beam irradiation, however the potential risks associated with it have not been fully recognized. Awareness of this problem is emphasized in \cite{problem-c-beam-irrad}, where the authors discuss this effect in steels.      

With the prospect of eliminating carbon from irradiated surfaces, an Oxygen Plasma Cleaning (OPC) technique has been explored on irradiated electrodes, aiming to remove the carbon contamination. 
Plasma cleaning is a versatile technique widely utilized across various industrial sectors, including semiconductors, biomedical applications, aerospace, and advanced material processing \cite{book,semiconductors,Biomedical, AdvMat}. This technique is particularly valuable where precision, cleanliness, and minimal surface damage are critical. Recently, plasma cleaning has been adopted for specialized applications in particle accelerators - oxygen plasma or mixtures of oxygen and argon have been employed to effectively remove hydrocarbon contamination from components such as beam line optics \cite{Moreno1,Moreno2}, accelerator vessels \cite{Giordano_2024}, copper cavities \cite{cleaning-copper-cavities} and superconducting accelerating cavities \cite{SLAC_plasma_cavities}, significantly improving vacuum operational stability. Beyond carbon and hydrocarbons decontamination, OPC induces other effects, such as the formation of superficial oxide layers \cite{FeO,CuO,CuO_lowP, carbon_STEM}. This is particularly impactful for materials where the oxide formation deriving from the oxygen plasma treatment has been reported to modify and often improve wettability, adhesion and dielectric properties \cite{Polyimide,electricalprop, steel}. Studies made in \cite{LACOSTE2002, LATIFI2014} report strong oxidation of stainless steel surfaces in the form of iron and chromium oxides, upon plasma-based ion implantation, without any degradation of the surface aspect. Chromium oxides via oxygen plasma treatments have been associated in \cite{WANG2021} with improvement of corrosion resistance while retaining a considerable electrical conductivity.
Also, \cite{SONMEZ2016} explores the effects of RF plasma treatments in stainless steel, observing that O$_2$ environments produce high valence oxide states along with increased thickness of oxide film and a roughness decrease in a nano scale. 

In this work, high voltage tests were conducted on plasma treated electrodes, with the aim of comparing results between plasma treated and non-treated surfaces, in terms of breakdown behavior and electric field holding performance. CuBe2, Cu-OFE and SS316LN electrodes were the materials selected to conduct the experiments as it will be discussed later on. 
Microscopic techniques using Scanning Electron Microscope (SEM) and Energy Dispersive Spectroscopy (EDS) were primary used to analyze all the material's surfaces during the different experiments. Moreover, X-ray Photoelectron Spectroscopy (XPS) was used to additionally explore possible changes in surface chemical composition.

\section{Methodology and Materials} \label{sec:methods}

\subsection{Tested Materials} 

Seven materials were studied as candidates for the manufacture of future Radio-Frequency-Quadrupoles. Among those materials, covered in \cite{paper1_catarina} and \cite{paper2-arxiv-catarina}, titanium grade 5 (Ti6Al4V), stainless steel AISI 316LN (SS316LN) and copper beryllium (CuBe2) were the three materials that have better responded to high voltage pulsing tests, being able to achieve high electric fields (above 100 MV/m) and presenting good resistance to sustain vacuum arcing.
Preliminary plasma cleaning tests were carried out on  existing copper beryllium (UNS type C17200 with 98 wt\% copper and 2\% beryllium). After the proven feasibility of the experiment, two materials were chosen for machining new electrodes, namely pure oxygen-free copper, UNS type C10100 (Cu-OFE), as it presents nowadays the most common material in terms of RFQ manufacturing, and stainless austenitic chromium-nickel-molybdenum steel with nitrogen addition (AISI 316LN), UNS type S31653, as it was the material with the highest electric field attained in the previously mentioned tests. In addition, when compared to the titanium or the copper alloys, SS316LN presents an easier machinability, a wider and more consistent supply availability and is more cost-effective, both in terms of material price and machining expenses.

\subsection{Irradiation and Breakdown Testing} 

Prior to oxygen plasma treatment, irradiation using low energy H$^-$ ions at 45 keV was performed in the cathode's surface for each material. For comparability, the irradiation has been set identically to the one reported in \cite{paper2-arxiv-catarina}.
In terms of the main chronology of the events, electrodes from the different materials start from being exposed to irradiation, followed by plasma treatment, and at last installed for breakdown testing on the Large Electrode System (LES) \cite{Classification_BD_anton, BD_statistics, KILDEMO2004596, LES_Anton_PhysRevAccelBeams, parallel_plates_dc_system}.
The LES aims to apply high pulsed DC fields  to a pair of electrodes (cathode and anode) with a small gap (60 $\mu$m) between the two surfaces. By controlling the voltage applied between anode and cathode, it is possible to monitor vacuum arcing triggered during the process. After sample installation and after a vacuum in the order of 10$^{-8}$ mbar is reached, the process of conditioning starts by applying small levels of voltage. From here, the voltage is gradually increased, in accordance with the material's response to the consequent breakdowns. Operational maximum field is typically  measured at a steady-state average breakdown rate of $10^{-5}$. The system allows for breakdowns real-time monitoring, along with positioning on the surface, while controlling voltage control. 
Additional information about the components and measurements techniques will not be described in this publication, as they have been addressed in detailed in \cite{paper2-arxiv-catarina}.

\subsection{Oxygen Plasma Treatment Set-up}

\begin{figure}[H]
    \centering
    \includegraphics*[width=0.9\columnwidth]{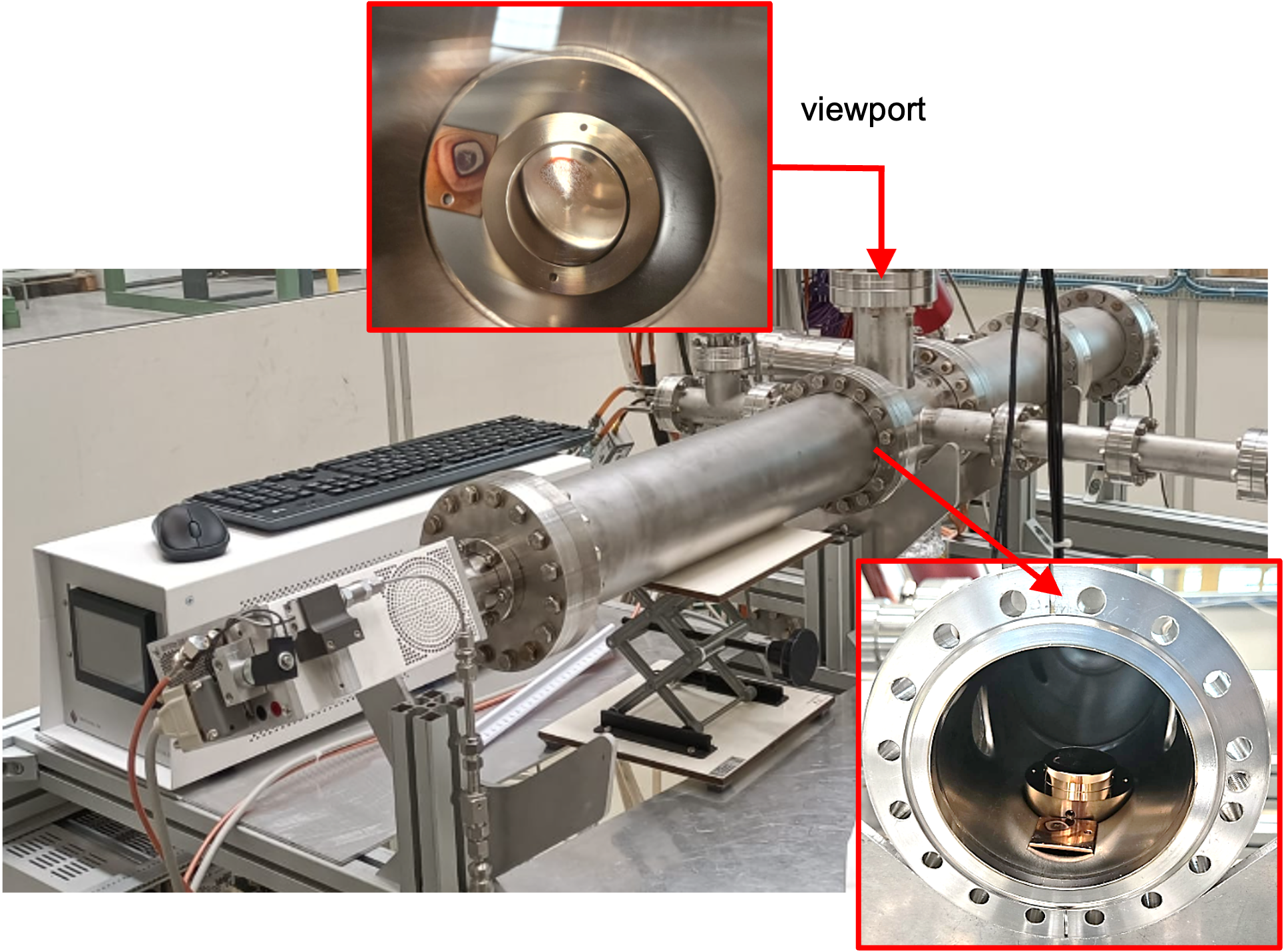}
    \caption{Photograph showing the main treatment chamber, in which an electrode is installed. The upper small photograph shows the viewport of the system, allowing visualization of the sample during operation. The lower small photograph shows the position of the CuBe2 electrode before closing the flange of the system.}
    \label{setup}
\end{figure}

The plasma cleaning treatments were performed in a dedicated test bench, which is shown in Figure \ref{setup}. 
The experimental set-up consists of the following components: a DN100 stainless steel vacuum vessel to house the treated samples, an oxygen-fed plasma source, and a turbomolecular pump mounted on a DN60 transition to maintain constant pressure. The plasma source employed is an inductively coupled plasma (ICP) RF sources branded ibss Group, operating at 13.56 MHz and an RF excitation power P$_\mathrm{RF}$ ranging from 10 to \SI{300}{W}. The gas, pure O$_{2}$ (99.999 $\%$), is injected in the ignition chamber of the plasma source and the active species are extracted through an orifice to the volume to be cleaned.  Completing the set-up are: a membrane gauge for the pressure reading, a quartz crystal microbalance coated with amorphous carbon as a reference for the etching rate, and a viewport for visual inspection of the samples.

During the treatments, the chamber pressure was kept at 4 $\times$10$^{-3}$ mbar. The other process parameters were adjusted based on the electrodes being treated. For the pilot test, the CuBe2 cathode was treated at 60 cm away from the plasma source, with plasma source powered at 50~W, and for a total treatment duration of 6 hours, split in tranches of 75, 105, and 180 minutes for intermediate inspections. 
Based on findings regarding carbon removal rates from the irradiated surfaces, subsequent treatments for the remaining two electrode pairs (Cu-OFE and SS316LN) were set to a fixed 6-hour duration. As precautionary measurement, the samples were placed at closer distance (30 cm) and the P$_\mathrm{RF}$ increased from 50~W to 100~W to ensure complete carbon removal. The plasma cleaning efficiency as function of chamber pressure, plasma source power and distance from the source outlet were the subject of a study that will be published elsewhere. In this work, the treatment parameters were selected to maximize the carbon removal rate while maintaining a conservative ion energy and minimize the physical sputtering of material from the sample and ion implantation.

\section{Results and Discussion} \label{sec:results} 
    \subsection{Plasma Treatment of a rough surface: CuBe2 electrodes}

Recent results, published in \cite{paper2-arxiv-catarina}, showed that for irradiated materials the capability of reaching high electric fields has been limited by the carbon content present at the cathode surface. Standard wet surface cleaning methods, using either solvents or detergents as it is common practice for UHV applications \cite{taborelli2020cleaningsurfaceproperties}, are not effective in removing amorphous carbon. Therefore, a new approach has been tested, by using OPC. This method would also have  the advantage that it could be applied in-situ on existing RFQ structures, in case carbon-contamination from irradiation would degrade operational performance.

CuBe2 has been irradiated and tested in the LES system. For this part of the study, the same irradiated CuBe2 electrodes were chosen to conduct a pilot test in carbon removal after LES testing using OPC.
In order to understand what would be the ideal time of treatment, without compromising the surface of the electrode, the procedure was divided as mentioned above in three treatments of 75 + 105 + 180 minutes.
Between each treatment, SEM and EDS analysis were performed to assess the surface and carbon state.

\begin{figure}[]
\includegraphics[width=0.46\textwidth]{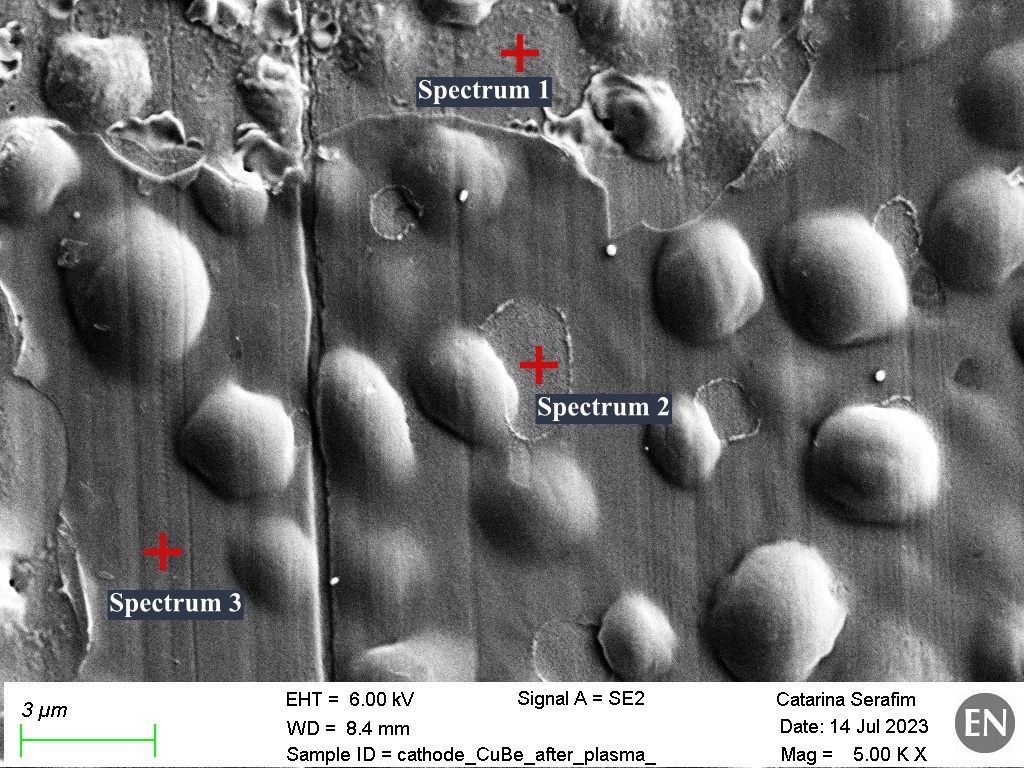}
\caption{\label{fig:1st_treatment_EDS} Scanning electron microscope imaging of the CuBe2 cathode's surface after the 1st plasma treatment, with identification of the location of the different EDS measurements points.  }
\end{figure}

After the first 75 minutes, SEM observation showed already that in some regions the CuBe2 surface was beginning to be exposed - see for example, regions marked as 'Spectrum 2' in Figure \ref{fig:1st_treatment_EDS}. 
An additional 105 minutes of treatment clearly showed that the carbon content has been reduced significantly, with the EDS values showing approximately a decrease of one third in comparison with the treatment before. 
However, these first two treatments, showed to be not sufficient in removing the carbon content in the CuBe2 surface.

\begin{table}[]
\small
\caption{Measurements of carbon content by EDS analysis made on CuBe2 cathode's surface, in different stages of the plasma treatment and on the locations visualized in Figs. \ref{fig:1st_treatment_EDS} and \ref{fig:3rd_tratment_EDS}. }
\label{table_EDS}
\begin{tabular}{llll}
 &  & C (wt\%) &  \\ \cline{1-3}
\multirow{3}{*}{\begin{tabular}[c]{@{}l@{}}75 minutes\\ of treatment\end{tabular}} & Spectrum 1 & 0.49 &  \\
 & Spectrum 2 & 3.46 &  \\
 & Spectrum 3 & 12.23 &  \\ \cline{1-3}
\multirow{7}{*}{\begin{tabular}[c]{@{}l@{}}360 minutes\\ of treatment\end{tabular}} & Spectrum 4 & 10.98 &  \\
 & Spectrum 5 & 4.02 &  \\
 & Spectrum 6 & 0.97 &  \\
 & Spectrum 7 & 0.10 &  \\
 & Spectrum 8 & 9.98 &  \\
 & Spectrum 9 & 3.16 &  \\
 & Spectrum 10 & 0.43 & 
\end{tabular}
\end{table}

\begin{figure}[]
\includegraphics[width=0.46\textwidth]{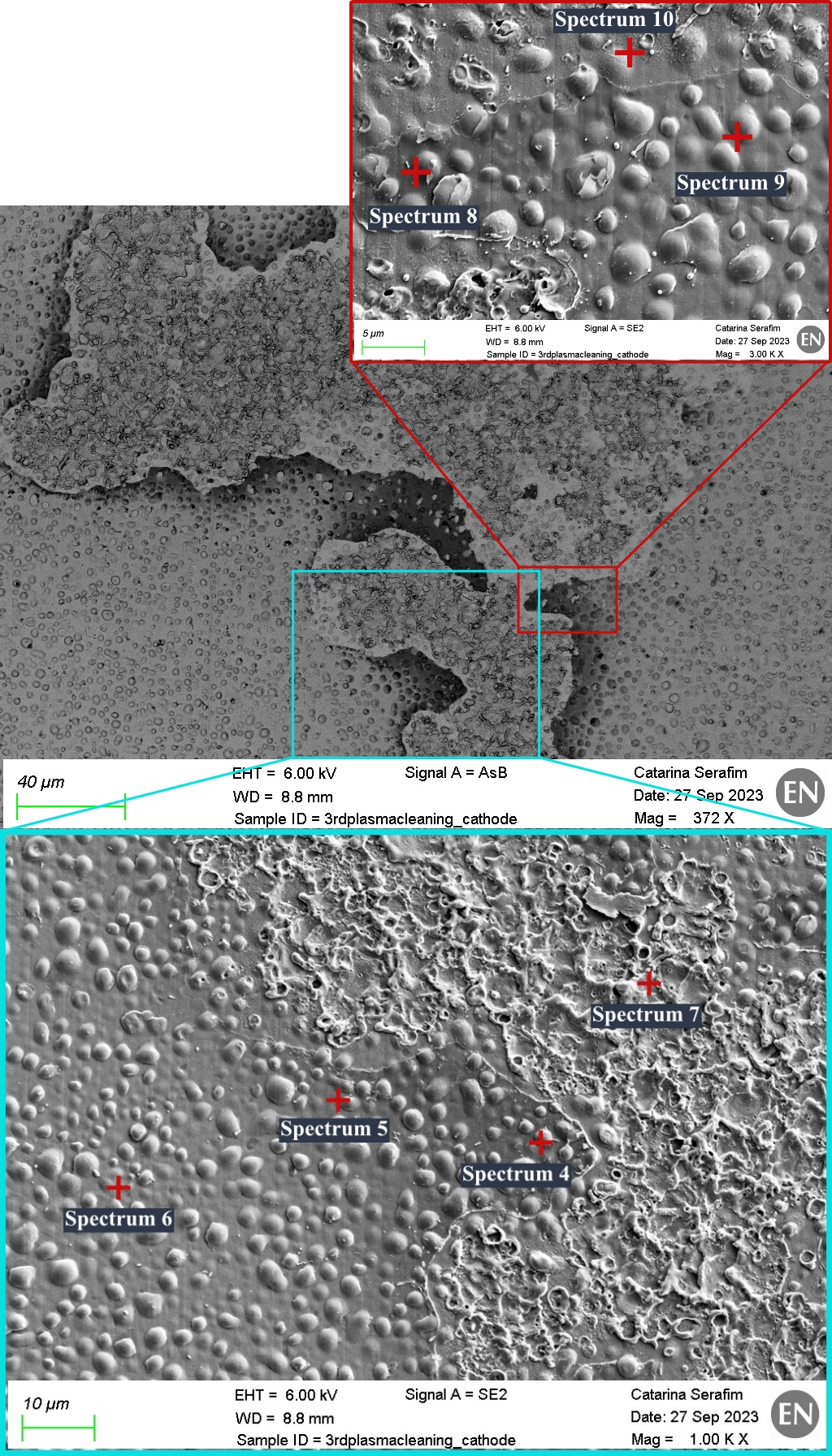}
\caption{\label{fig:3rd_tratment_EDS} Scanning electron microscope imaging of the CuBe2 cathode's surface after 360 minutes of plasma treatment, with identification of the location of the different EDS measurements points. }
\end{figure}

Figure \ref{fig:3rd_tratment_EDS} shows the surface state after a final treatment with additional 180 minutes. The main image, represents a region with higher blister concentration, where breakdowns from LES experiment can also be seen. The image was acquired using backscattered detector, providing information about the locations with existing carbon in a darker coloration. EDS measurements were done in multiple locations (blue and red images). The carbon layer has proven to be difficult to be completely removed from the regions contouring the edges of the pre-existing breakdowns craters,  possibly due to copper vaporization during vacuum arcing events. 
Apart from these specific regions, the full treatment proved to be successful in removing the carbon, with final concentrations staying below 1 wt$\%$.

Taking in consideration, that the plasma treatment method was aimed to be used further in our studies on surfaces free from breakdowns and minimal roughness, 360 minutes (6 hours) was the time chosen for the duration of treatment for future electrodes.

\subsubsection{Re-conditioning of CuBe2 electrodes after plasma treatment }

\begin{figure}[H]
\includegraphics[width= 0.48\textwidth]{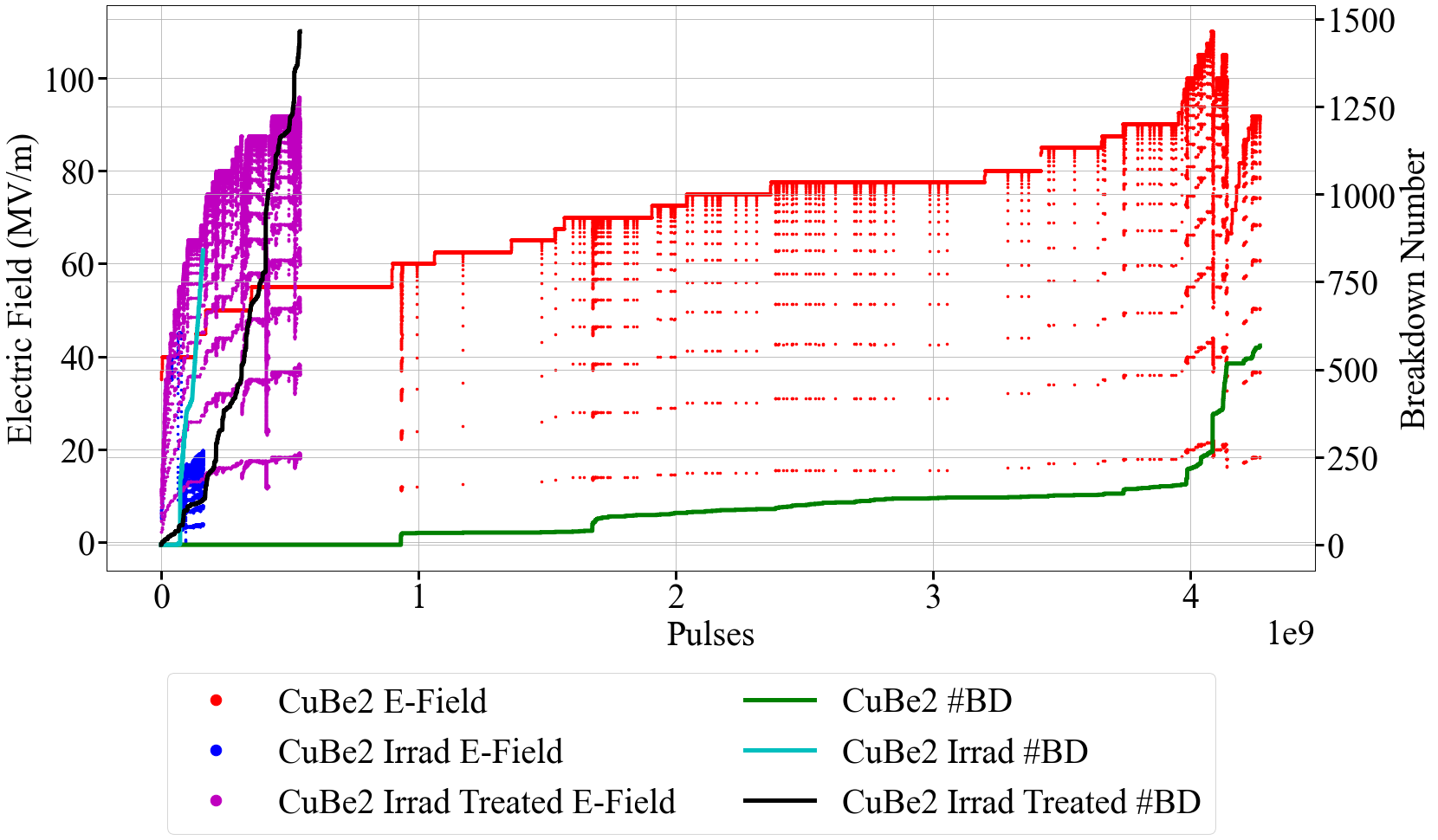}
\caption{\label{fig:CuBe2_conditioning} Overlap of the different conditioning plots of LES tests on CuBe2 electrodes (non-irradiated, irradiated and irradiated and plasma treated).}
\end{figure}

Figure \ref{fig:CuBe2_conditioning} represents the results from the different LES testing, made in CuBe2 electrodes. The explanation regarding the different conditioning curves can be found in subsection \ref{conditioning_results_section_explained}.
The plot shows the conditioning of CuBe2 irradiated (pre-plasma treatment) and non-irradiated electrodes from publication \cite{paper2-arxiv-catarina}, plotted simultaneously, with the new result of the conditioning test made with the plasma treated CuBe2 in magenta.
Again, it should be emphasized that this is the same pair of electrodes that was previously exposed to high voltage pulsing test after irradiation (dark blue curve). 
The results from the new conditioning show a dramatic difference in terms of higher electric field achieved, by being able to reach a maximum of 95 MV/m and a stable field of 92 MV/m, almost 6x more than the irradiated electrodes without plasma treatment (approximately 17 MV/m).
This result stays however slightly below the electric field reached by the non-irradiated electrodes (red curve), where the maximum field was 110 MV/m.

\begin{figure}[]
\centering
\includegraphics[width=0.45\textwidth]{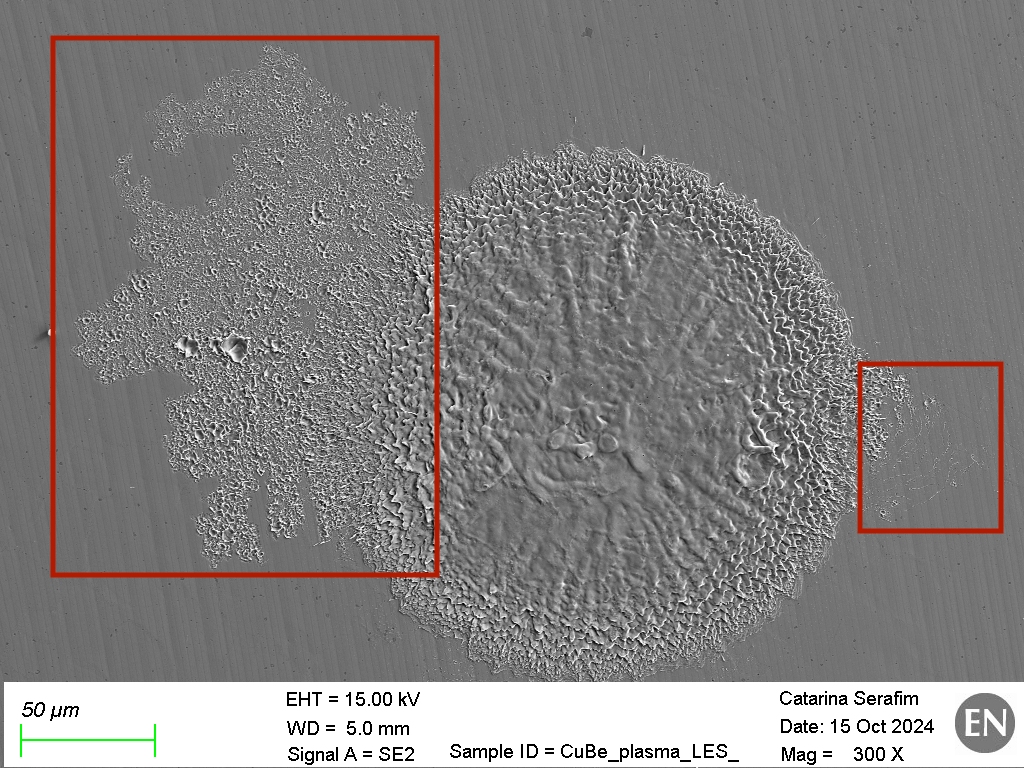}
\caption{\label{fig:BD-2conditionings-cube} SEM image using secondary electron detector with 300x magnification showing small breakdowns caused by the 1st conditioning test (inside the red squares) and a single bigger breakdown (event during the 2nd conditioning test).  }
\end{figure}

The small difference is possibly attributed to the carbon content that was not removed in some of the regions on the cathode surface. Additionally, the roughness of the surface in both cathode and anode, due to the breakdowns from the previous test, may have contributed to a more hard process of conditioning, which could also explain the higher number of breakdown events for the irradiated and treated electrodes when compared with the total number of breakdowns from the non-irradiated electrodes test.

\begin{figure}[]
\includegraphics[width=0.47\textwidth]{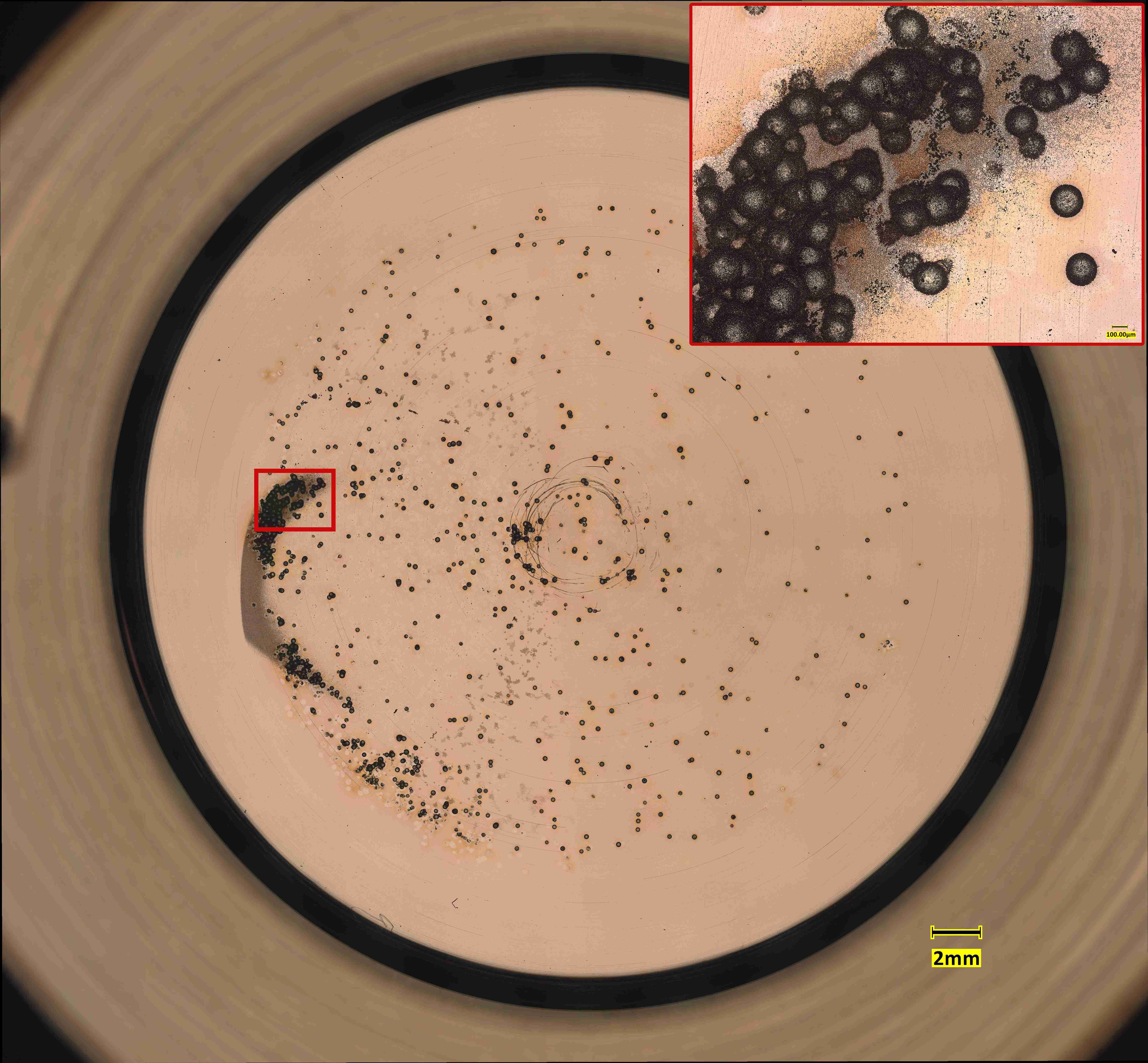}
\caption{\label{fig:optical-cube2-mapping} Optical image of CuBe2 cathode surface and zoom in region in red, after retest on the LES system (2nd conditioning). }
\end{figure}

The smaller vacuum arcs from the first conditioning can be observed inside the red squares in Figure \ref{fig:BD-2conditionings-cube} in contrast with a breakdown from the 2nd conditioning, where a single breakdown reached diameters in the order of 300 $\mu$m. This difference in size is due to the lower energy discharged at lower fields, see \cite{energy_dependence_BD}. Figure \ref{fig:optical-cube2-mapping} shows the mapping of the surface using an optical microscope, in which the C-shape mark on the left, visible on the surface, is the optical effect caused by the blistering (irradiated region). The small black dots dispersed in the surface correspond to the breakdowns from the 2nd conditioning. For comparison of the previous surface states, please see Figure 11 from \cite{paper2-arxiv-catarina}. 

While a good dispersion of breakdowns in the surface is observed, a few clusters have been identified in the irradiated region. These breakdown clusters, that were formed during the second conditioning experiment, are coincident with the surface regions in which carbon was not removed from the plasma treatment (as seen in Figure \ref{fig:3rd_tratment_EDS}). 
In the second conditioning, the presence of traces of carbon seemed to attract vacuum arcing to those specific locations, leading to an increase of more than 2x the total number of breakdowns when compared with the non-irradiated electrodes.

\subsection{Plasma treatment of pristine irradiated surfaces: Copper and Stainless Steel}

New electrodes of Cu-OFE and SS316LN were machined, irradiated and submitted to a treatment of 6 hours of OPC.
For copper, the carbon layer present on the electrode surface after irradiation is very clearly observable when using SEM. In some regions, this carbon layer has been broken, showing some flakes partially peeling from the copper surface - the effect is visible in Figure \ref{SEM_Cu_before_plasma_figs}. 

\begin{figure}[H]
\centering
\subfigure[]{\includegraphics[width=0.47\textwidth]{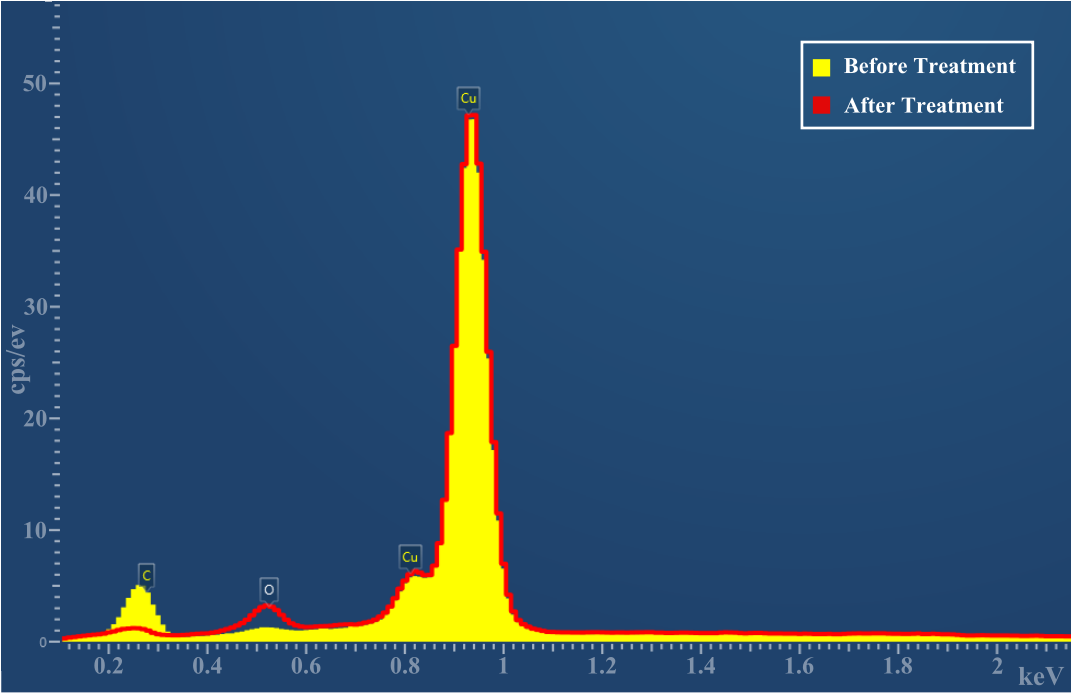}}
\subfigure[]{\includegraphics[width=0.47\textwidth]{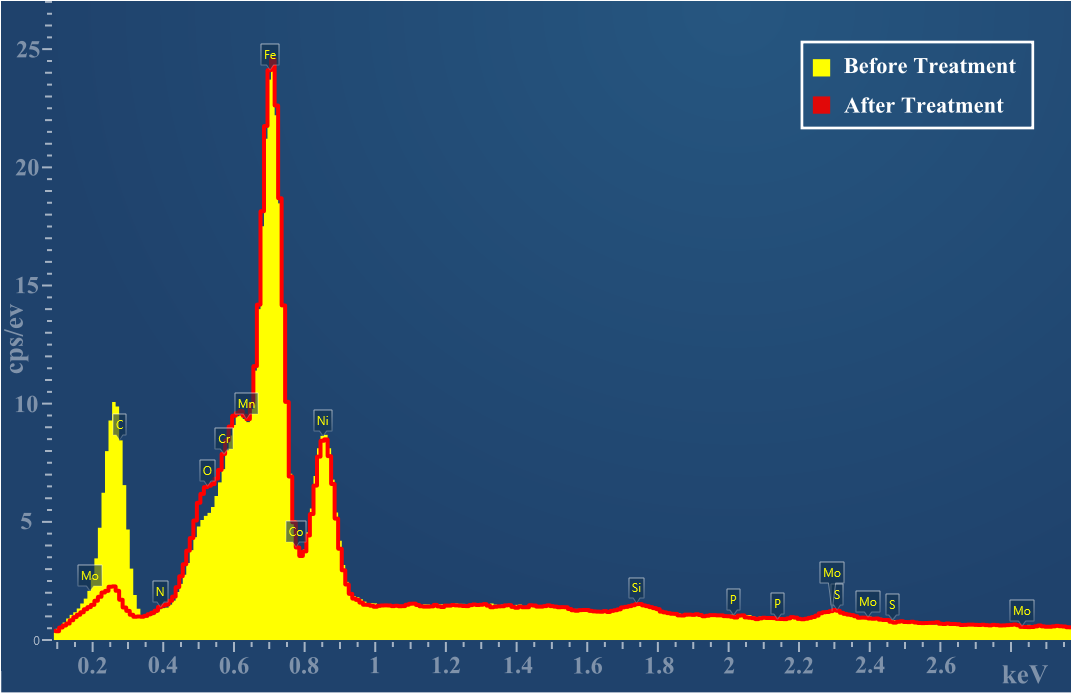}}
    \caption{EDS spectrum analysis for (a) Cu-OFE and (b) SS316LN, showing the elements present in the surface before plasma treatment (yellow) and after plasma treatment (red). }
    \label{EDS_spectrum_copper_ss}
\end{figure}

\begin{figure*}[]
\includegraphics[width=0.328\textwidth]{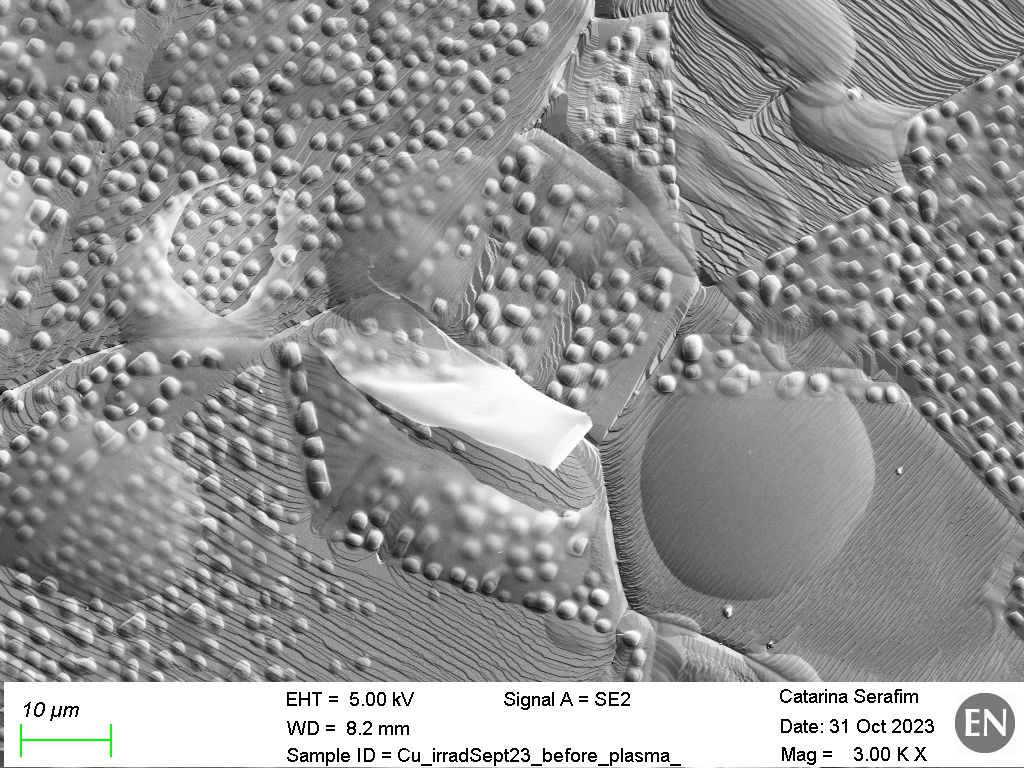}
\includegraphics[width=0.328\textwidth]{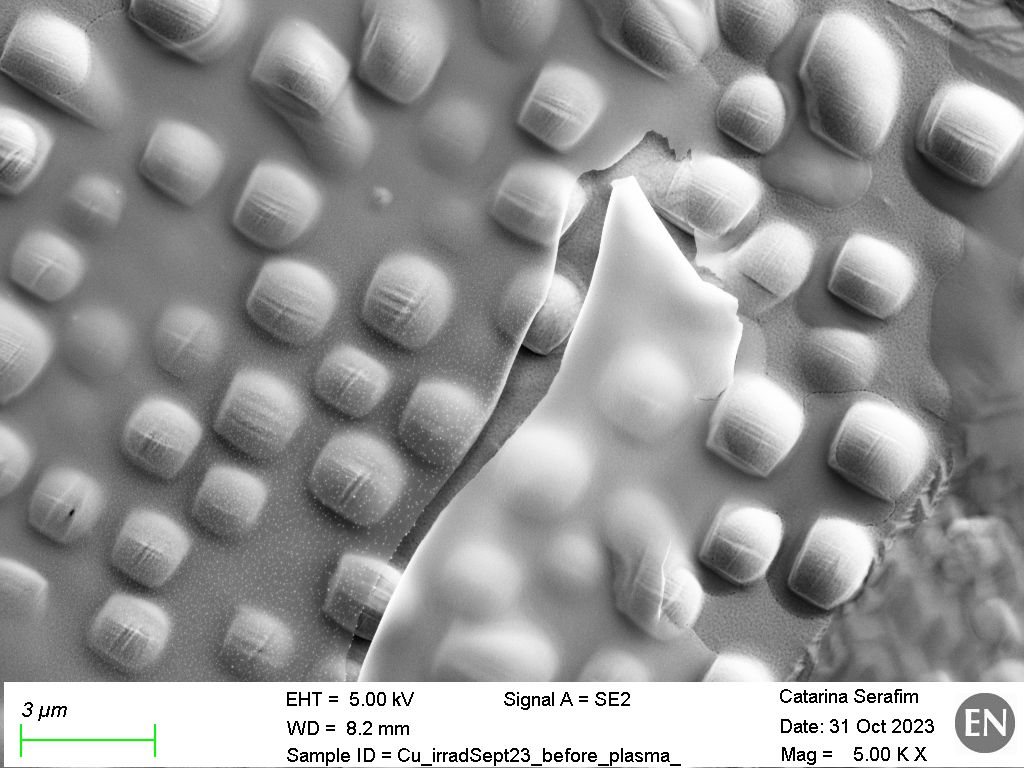}
\includegraphics[width=0.328\textwidth]{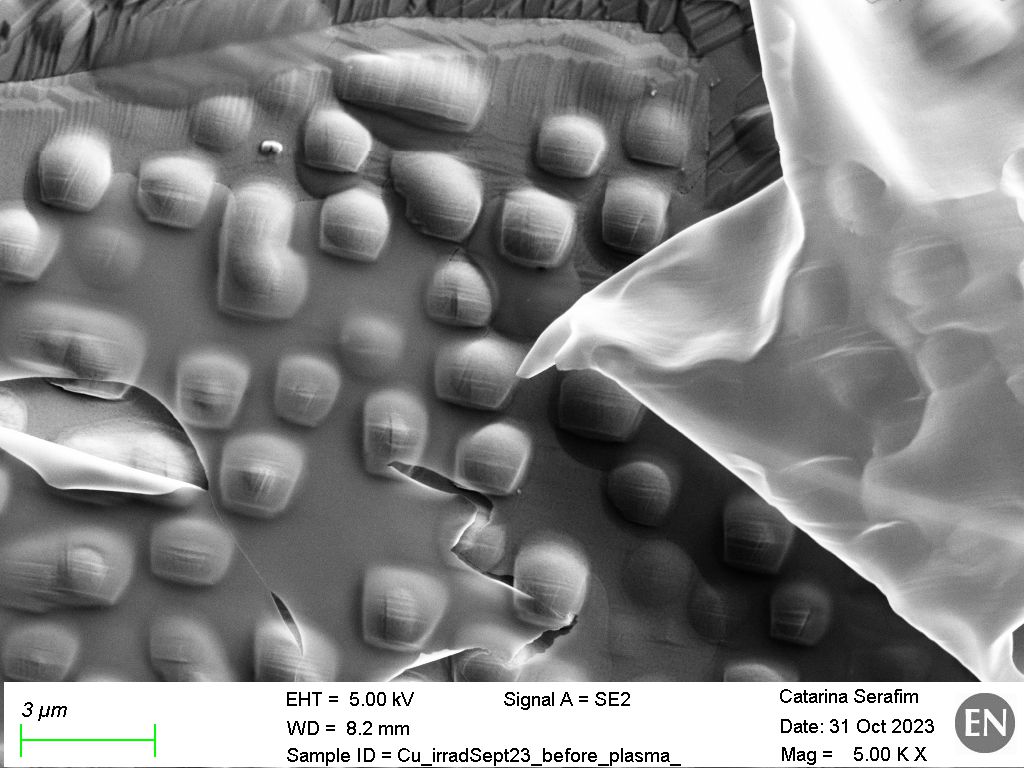}
    \caption{SEM images using secondary electron detector showing the state of the surface of Cu-OFE electrode after irradiation and before OPC. In the images, the blistering caused by the H implantation in Cu-OFE is visible, additionally the formation of a thin carbon layer upon the surface is clear, which has been broken and is laminating from the surface. Also in the left image, some bubbles are visible, an effect caused by the implantation of hydrogen between the copper surface and the carbon layer.}
    \label{SEM_Cu_before_plasma_figs}
\end{figure*}

\begin{SCfigure*}[]
\includegraphics[width=0.35\textwidth]{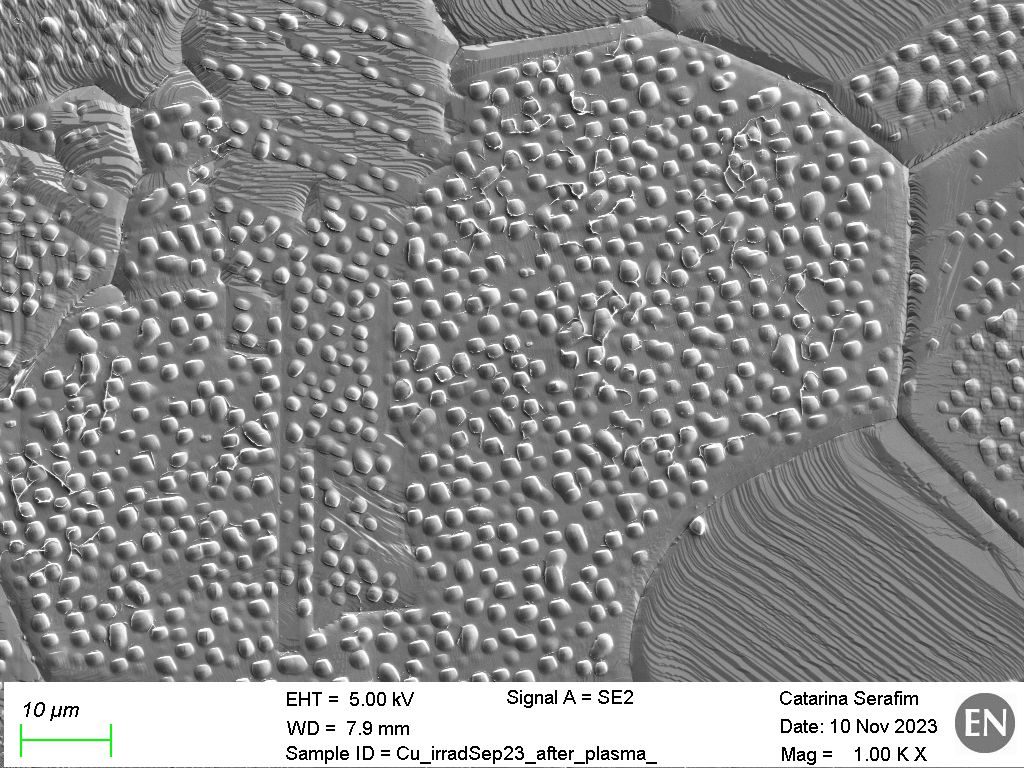}
\includegraphics[width=0.35\textwidth]{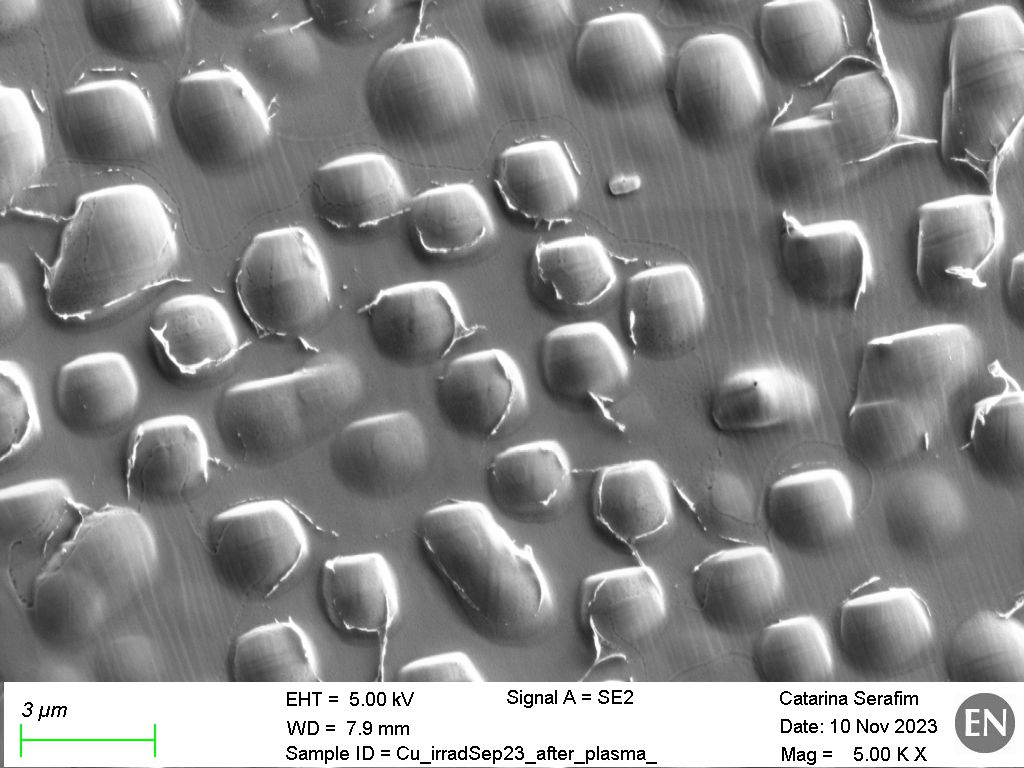}
   \caption{SEM images using secondary electron detector showing the state of the surface of Cu-OFE electrode after irradiation and after oxygen plasma cleaning. After OPC, bubbles on the surface are no longer visible and the continuous layer of carbon was removed, however, some carbon residuals are still visible at high magnification (right image). }
    \label{SEM_Cu_after_plasma_figs}
\end{SCfigure*}

In SS316LN, the presence of the carbon contamination doesn't create a distinguished carbon layer, as seen in copper and copper beryllium. SEM and optical microscopic analysis show no indication of a layer formation above the surface, however EDS measurements in multiple regions inside the irradiated region confirmed the same percentages of carbon as on copper -  between 8-12 wt$\%$.
EDS analysis were done after irradiation,  before and after plasma cleaning. The treatment proved to be very effective in removing the carbon, where EDS analysis in different locations consistently measured carbon below 1 wt$\%$ for both materials. 

In Cu-OFE some residuals of the carbon layer were still visible on the copper surface, see Figure \ref{SEM_Cu_after_plasma_figs}. According to EDS analysis, these showed no significant carbon contamination, while prolonging the treatment to further cleanliness could have had increased the surface oxidation on the surface. 
Figure \ref{EDS_spectrum_copper_ss} shows the EDS elements spectrum for (a) Copper and (b) SS316LN, before and after plasma cleaning, where quantitative differences in the carbon peak are visible. After plasma treatment, an increase of oxygen content in both samples is also evident, which confirms the oxidative nature of the treatment.

    \subsubsection{Conditioning Results \label{conditioning_results_section_explained}} 
The plots in Figures \ref{fig:Cu_conditioning} and \ref{fig:SS316_conditioning} include three different set of conditioning tests, for three different pairs of electrodes.
The conditioning curves represent the continuous increase of voltage for one pair of electrodes, which is translated in electric field (left vertical axis). 
The breakdowns during the process are visible via sudden drops in electric field, followed by a rapid recovery to voltage levels previous to the breakdown. From this point the voltage is kept during a certain number of pulses until a new increase of voltage is set. 
The conditioning curves are represented in red (non-irradiated electrodes), dark blue (irradiated electrodes) and magenta (irradiated and plasma treated electrodes).
The respective cumulative number of breakdowns during the conditioning are represented in green, cyan and black and can be read through the right vertical axis.

\begin{figure*}[!tbh]
\centering
\subfigure[]{\includegraphics[width=0.32\textwidth]{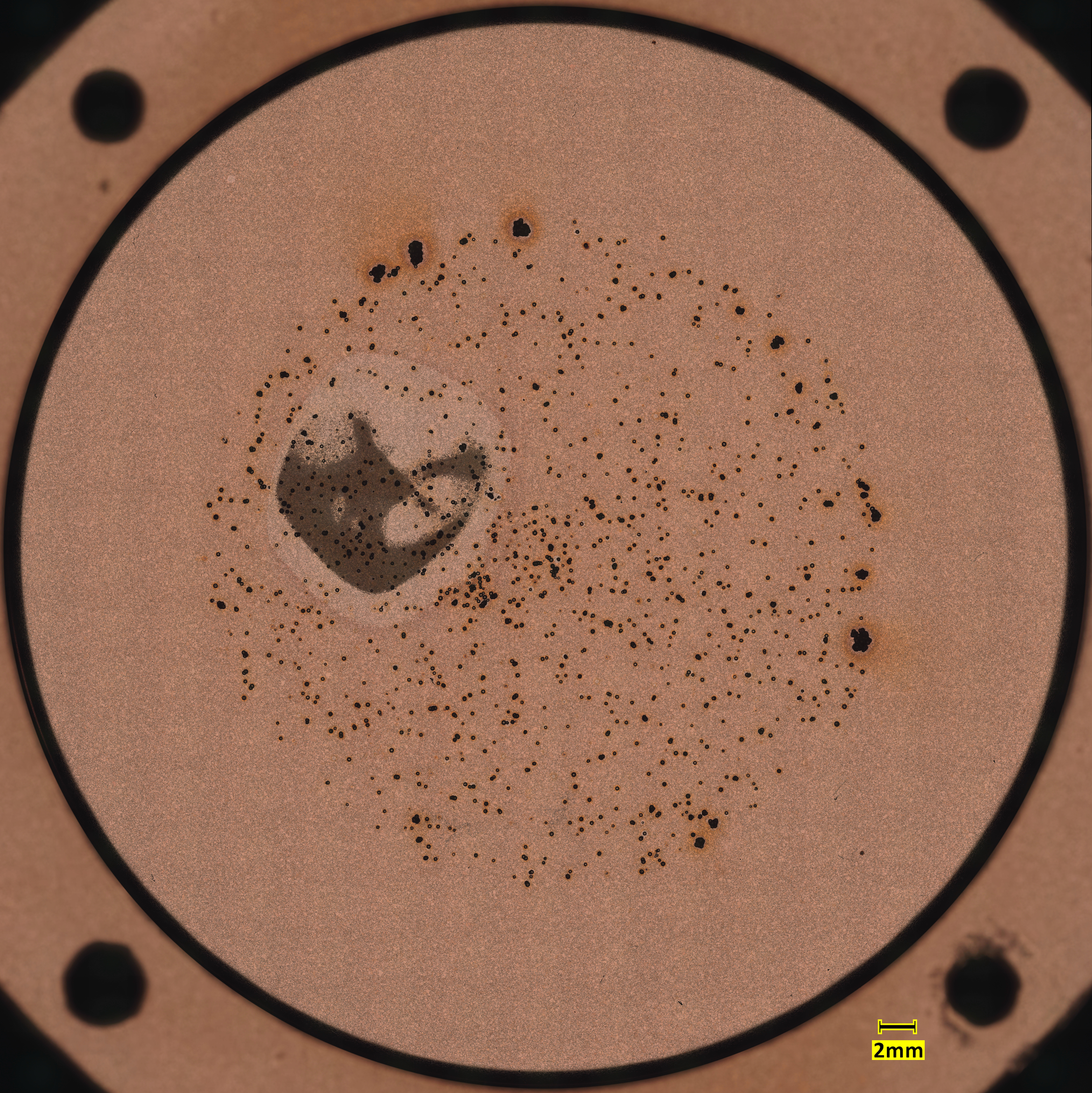}}
\subfigure[]{\includegraphics[width=0.323\textwidth]{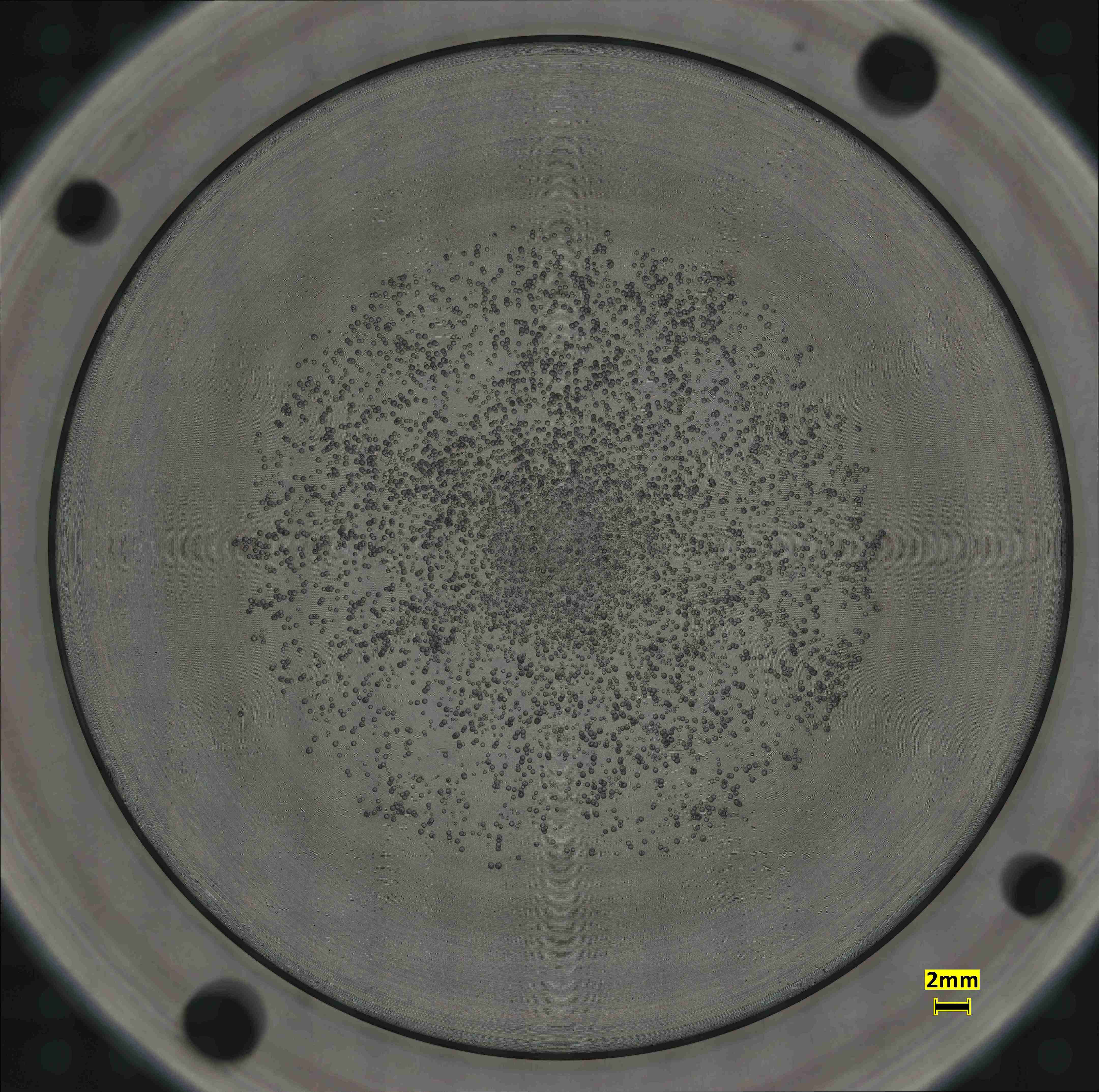}}
\subfigure[]{\includegraphics[width=0.338\textwidth]{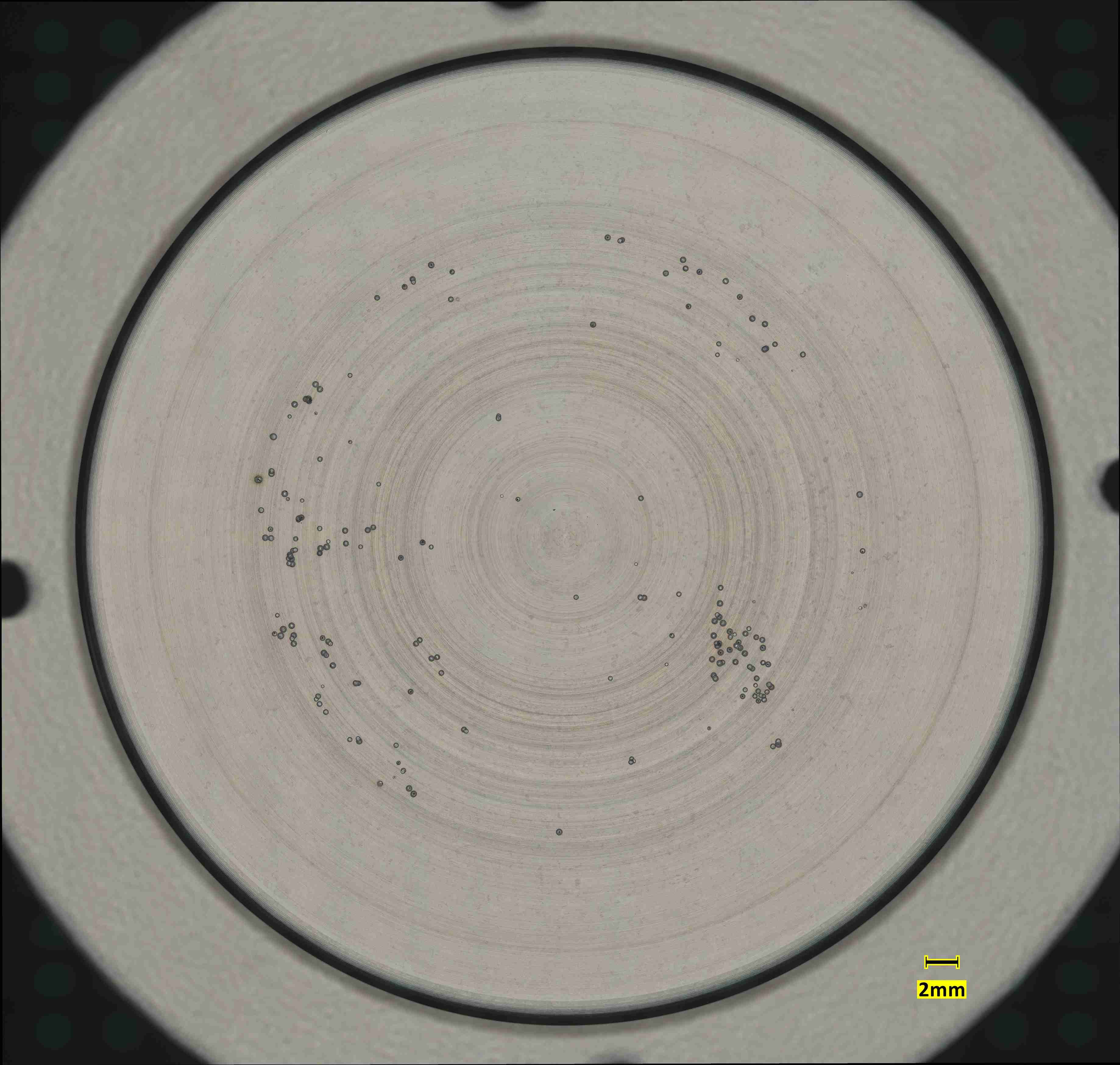}}
    \caption{Optical image of three different cathodes after LES testing: (a) irradiated and plasma treated Cu-OFE cathode surface, (b) non-irradiated SS316LN cathode surface, and (c) irradiated and plasma treated SS316LN cathode surface. \label{optical-figures-ss}}
\end{figure*}

        \subsubsection*{Copper (Cu-OFE)}

\begin{figure}[h]
\includegraphics[width=0.48\textwidth]{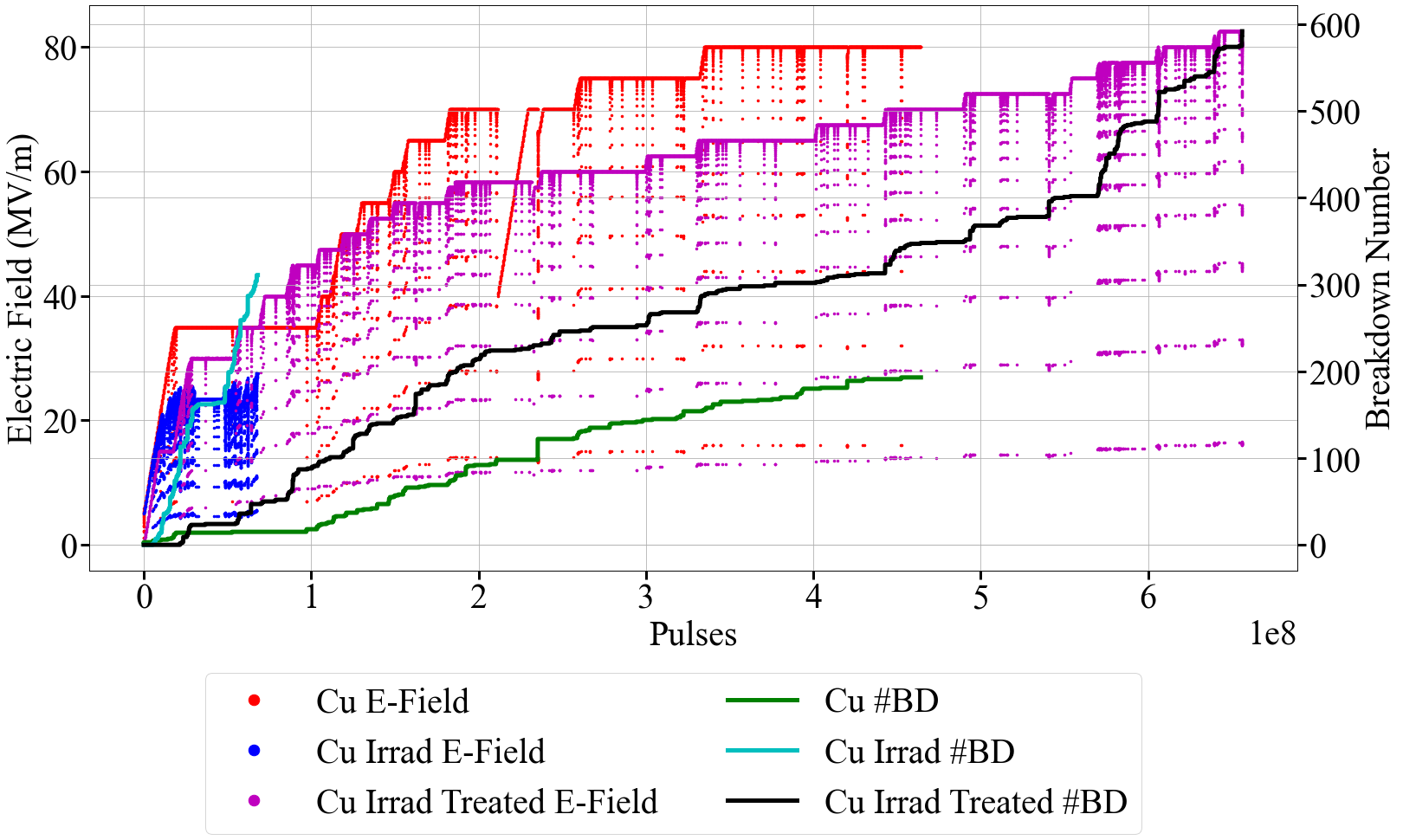}
\caption{\label{fig:Cu_conditioning} Overlap of the different testing conditioning made in three different Cu-OFE electrodes (non-irradiated, irradiated and irradiated and plasma treated). }
\end{figure}

\begin{figure}[H]
\includegraphics[width=0.48\textwidth]{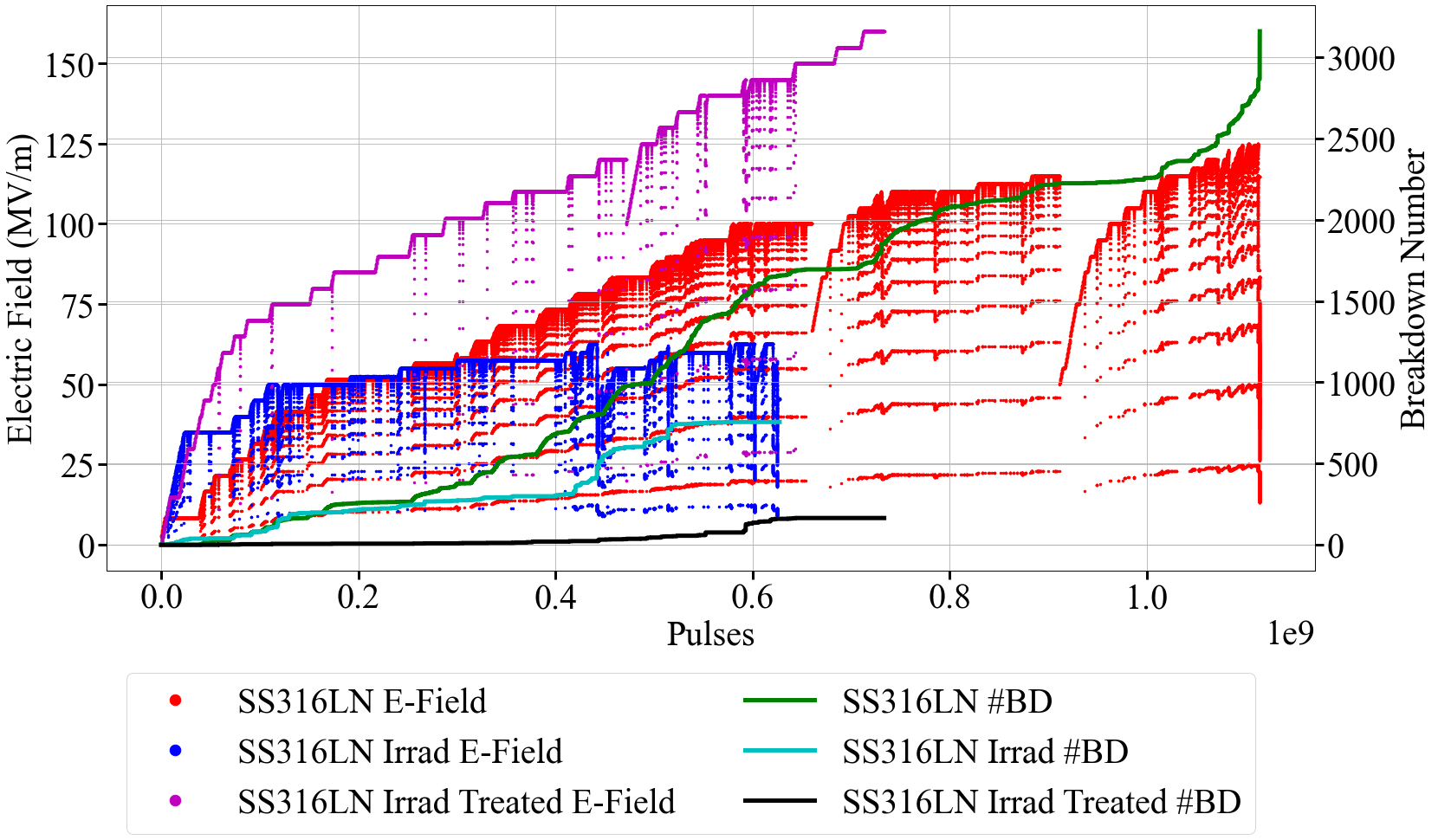}
\caption{\label{fig:SS316_conditioning} Overlap of the different testing conditioning made in three different SS316LN electrodes (non-irradiated, irradiated and irradiated and plasma treated). }
\end{figure}

Figure \ref{fig:Cu_conditioning} shows simultaneously the LES testing results for three different Cu-OFE electrodes. 
The conditioning of Cu-OFE irradiated and treated electrodes is shown in the red curve in Figure \ref{fig:Cu_conditioning}.
Both Cu non-irradiated and irradiated plasma treated have reached the mark of 80~MV/m.

When tested with high pulsing voltages in the LES, irradiated electrodes (without any plasma treatment) have only stably reached 23~MV/m - dark blue curve in Figure \ref{fig:Cu_conditioning}. For the latter, surface analysis have shown breakdown to be positioned in the irradiated region, following the same behavior as the CuBe2 irradiated electrodes reported in \cite{paper2-arxiv-catarina}. 

The irradiation didn't provoke sudden and agglomeration of vacuum arcs in specific locations or any decrease in performance when compared with the non-irradiated pair (results as well as surface imagining are very similar to the ones reported in \cite{paper1_catarina}). The final state of breakdown dispersion is shown in Figure \ref{optical-figures-ss} (a): the distribution of breakdowns is quite uniform all over the surface, showing no extraordinary events on the blistered irradiated region. Some small clusters have been observed on the edges of the tested surface, however apart from the concentration of multiple breakdowns, no additional features have been found.

        \subsubsection*{Stainless Steel (SS316LN)}
        
SS316LN non-irradiated electrodes were among the studied materials in \cite{paper2-arxiv-catarina}, being the material that presented the best results in terms of conditioning performance. The irradiated pair, however, has shown to being able to achieve only 60 MV/m (dark blue curve).
For the case of the SS316LN electrodes irradiated and plasma treated, the conditioning results, have not only reached the maximum electric field mark of the non-irradiated electrodes (120 MV/m - red curve), but have surpassed it, by reaching a maximum and constant field of 160 MV/m (magenta curve).
The irradiated and treated electrodes have registered until the end of testing less than 200 breakdowns (black curve). It is the first time in breakdown testing that, for this level of electric fields, such a small number of total breakdowns is registered. For comparison, the non-irradiated electrodes have registered more than 3000 breakdown events.

Figure \ref{optical-figures-ss} includes optical micrographs of the status after high voltage testing for both cases: (b) non-irradiated and (c) irradiated, where the notorious difference in breakdown on the surface are visible. Irradiation on SS316LN material, have produced no blistering seen in the surface. Dark coloration has been observed after irradiation, and associated with the carbon content. After plasma treatment, the irradiated spot was no longer visible on the surface, leaving no traces of irradiation.
Additionally, after high voltage testing, SEM imaging showed no defects on the entire surface apart from the breakdowns. The latest, have reached large round crater diameters in the order of 500 $\mu m$, due to the large fields experienced during the testing.

The difference in both electric field performance and breakdowns events suggests that the plasma treatment, aimed to remove the carbon contamination from the stainless steel cathode, has produced an additional effect on the electrode's surface. Additional studies on the effects of OPC in SS316LN surface were therefore performed, as discussed in the next section.

\subsection{Analysis of chemical composition on plasma-treated SS316LN samples} \label{ss316-xps}

To explore what the implications of the oxygen plasma treatment on the irradiated surfaces are, three 40$\times$40 \SI{}{mm^2} SS316LN samples were prepared for XPS analysis. Sample \textit{i)} was partially masked and then irradiated using the irradiation test stand with the same parameters used for the electrodes described above. 

\begin{figure}[H]
\subfigure[]{\includegraphics[width=0.117\textwidth]{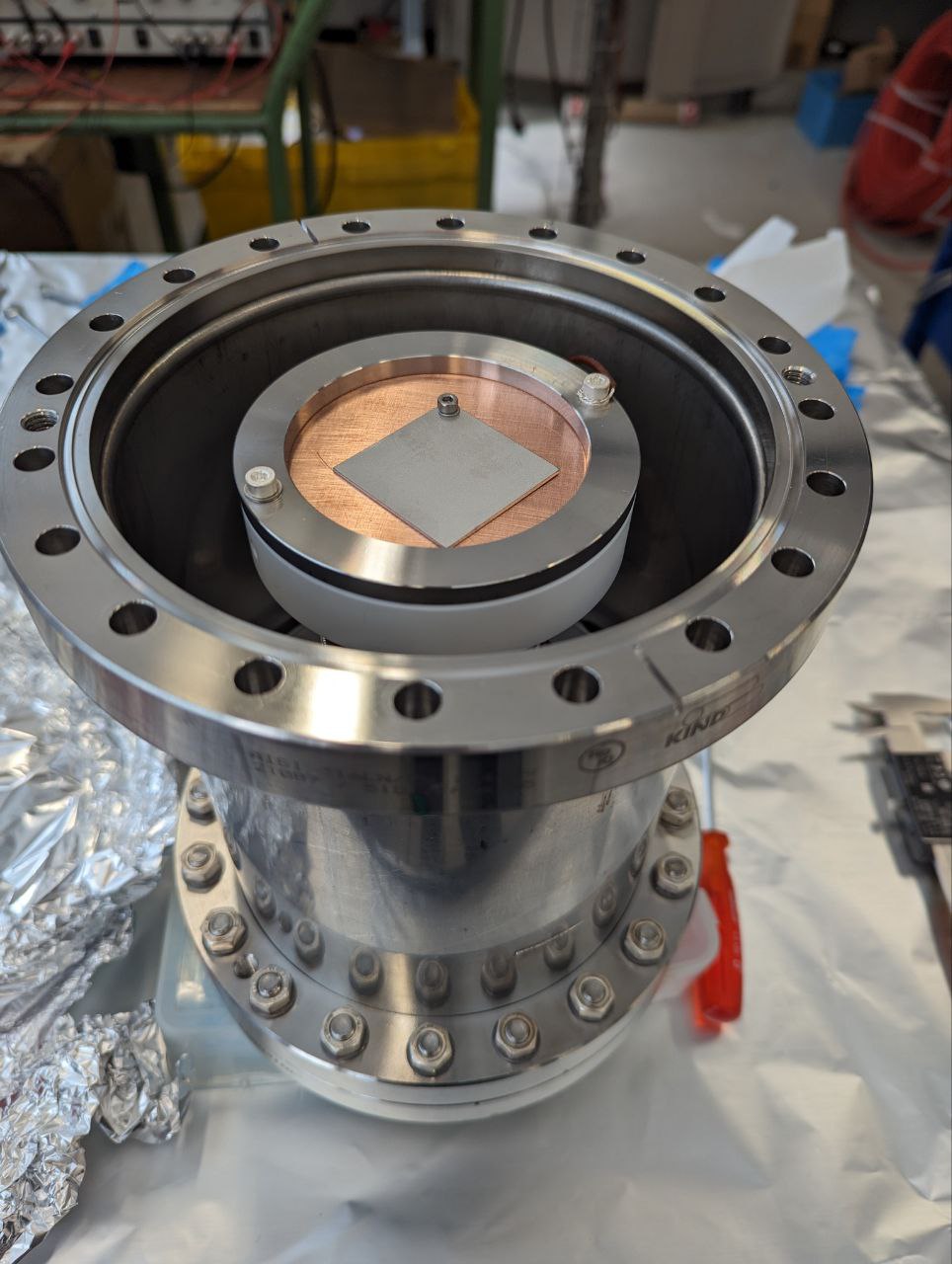}}
\subfigure[]{\includegraphics[width=0.155\textwidth]{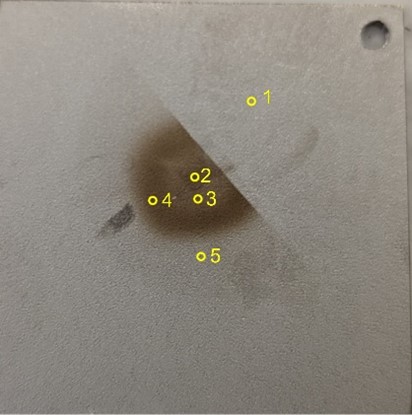}}
\subfigure[]{\includegraphics[width=0.197\textwidth]{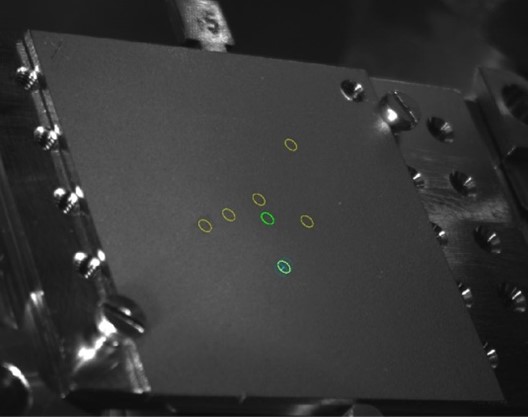}}
    \caption{(a) SS316LN sample \textit{i)} mounted onto the target chamber before installation chamber and irradiation. (b) Photograph of sample \textit{i)} after H- irradiation and before the plasma cleaning, including markings of regions where XPS analysis was performed. (c) Sample \textit{i)} after plasma treatment, mounted inside XPS chamber with identification of the measurement points.   }
    \label{sample_irrad_figures}
\end{figure}
Figure \ref{sample_irrad_figures} (a) shows the target sample mounted in the irradiation chamber. After confirmation of the presence of a carbon layer as found on the electrodes by EDS, the sample underwent the same oxygen plasma cleaning treatment as before. Figure \ref{sample_irrad_figures} (b) shows a photograph of the sample \textit{i)} after irradiation. The beam-exposed region is visible by its dark coloration. Figure \ref{sample_irrad_figures} (c) shows the same sample inside XPS chamber after the plasma treatment, where the irradiation marks are no longer visible. For comparison, two additional samples had been prepared and characterized in parallel. Sample \textit{ii)} was not irradiated, but treated by OPC after degreasing. Additionally, sample \textit{iii)}, called reference sample, was not submitted to any irradiation and/or treatment other than degreasing and was stored under primary vacuum until the XPS analysis was performed.

\begin{figure}[]
\includegraphics[width=0.48\textwidth]{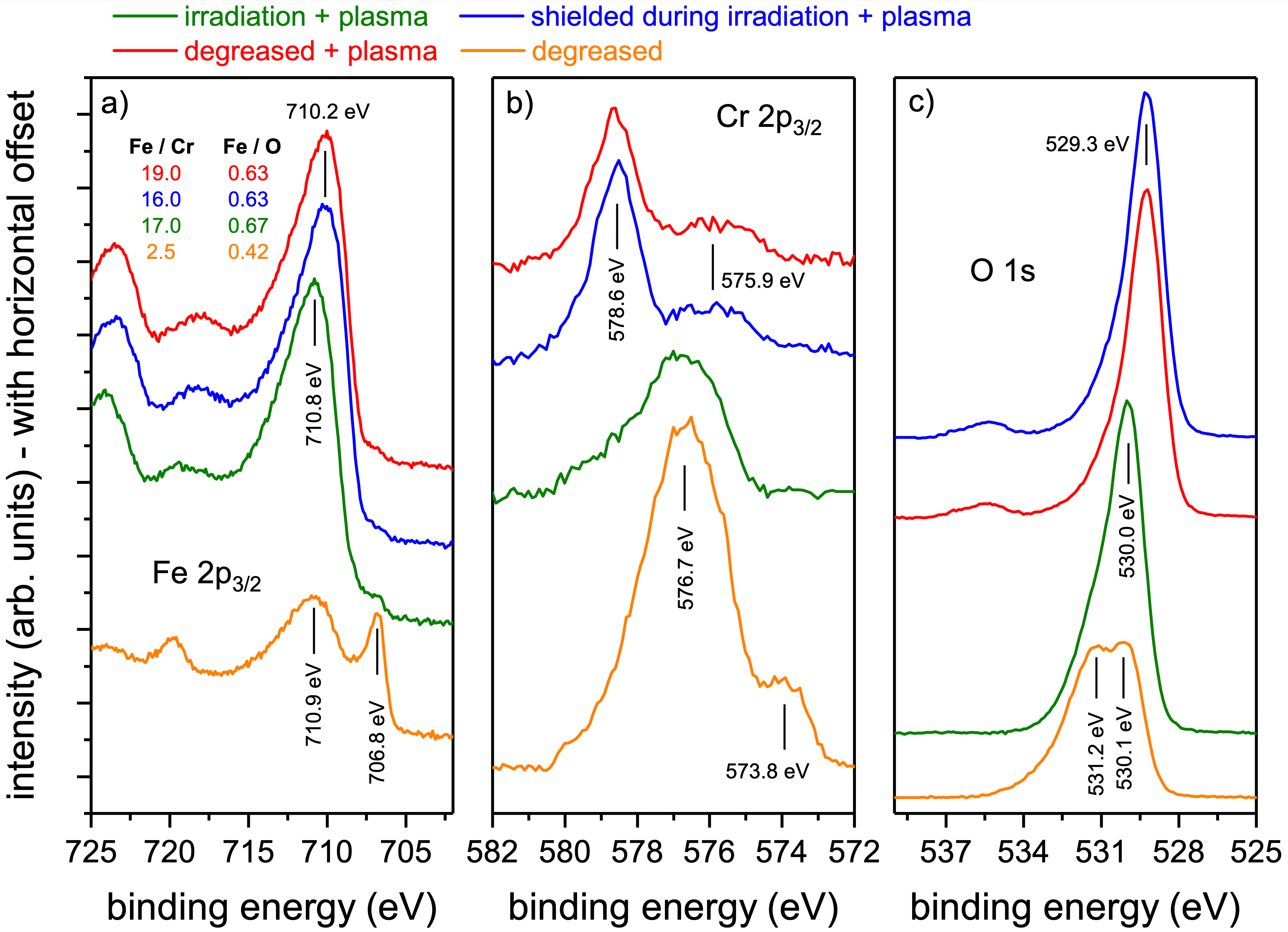}
\caption{\label{fig:xps-results} X-ray photoelectron spectra of SS316LN after different surface preparation procedures as indicated in the legend -- (a) Fe\,2p$_{3/2}$, (b) Cr\,2p$_{3/2}$, and (c) O\,1s state, respectively. The inset in a) shows the elemental ratios of Fe/Cr and Fe/O.}
\end{figure}

Figure \ref{fig:xps-results} includes the main core level spectra of the most relevant elements on the surfaces: iron (Fe\,2p$_{3/2}$), chromium (Cr\,2p$_{3/2}$) and oxygen (O\,1s) from the XPS characterization of the three samples. The green and blue spectra represent different measurements on the irradiated sample after plasma cleaning, in the irradiated and non-irradiated region, respectively. The red data represents the degreased and oxygen plasma treated sample \textit{ii)}, and the orange spectra are representative for the sample that was only degreased (without irradiation nor additional treatments). We have calculated the elemental composition assuming a homogeneous distribution of all elements by ﬁtting all spectra using CasaXPS version 2.3.24 \cite{FAIRLEY2021100112} with the algorithms implemented therein considering the electron analyzer transmission function. For all spectra, a Shirley-type background correction was performed, and the elemental sensitivity factors that are based on the cross-sections reported by Scoﬁeld \cite{SCOFIELD1976129} and taking into account an inelastic mean free path (IMFP) of the emitted electrons, which depends on their kinetic energy: IMFP proportional to E$_{kin}^{0.7414}$ \cite{Jablonski1993}, are considered.

Most significantly, the XPS results confirm the removal of carbon induced by the reactive oxygen species in the plasma \cite{Giordano_2024} as shown based on the EDS analysis. The C surface content of the degreased sample was 25\,at.\% and typical values for irradiated samples with carbon layer are above 90\,at.\%. The plasma treatment reduced the amount of the surface carbon to below 10\,at.\%. As the samples had to be transferred through air prior to surface characterization, such trace amounts are not avoidable due to readsorption of hydrocarbon molecules from ambient air. Figure\,\ref{fig:xps-results}\, (a) includes the Fe/Cr and Fe/O ratios of all four sample preparation stages. It is evident that the plasma treatment not only leads to further oxygen uptake and oxidation but also strongly depletes the chromium content at the surface, which is typically high for stainless steel samples. This change could be linked to a preferred oxidation of iron leading to redistribution of Cr to subsurface regions, or due to the possibility of creation of oxygen ions in the source that could sputter the outermost surface of the sample.

When analyzing the spectral signatures of the samples, it is clearly visible that the surface is stronger oxidized after the plasma treatment. The Fe\,2p$_{3/2}$ state at 706.8\,eV, which is clearly present on the degreased surface and is linked to metallic Fe$^0$ species at the surface \cite{BIESINGER20112717}, is almost absent for all other samples. Instead, the main contribution is from iron oxide (with an intensity maximum at 710.2\,-\,710.9\,eV). The spectral shape resembles the spectrum of Fe$_2$O$_3$ in Ref.\,\cite{Hughes2024}. One has to mention that the binding energy values of the core levels for the irradiated samples after plasma cleaning are 0.6-0.7\,eV higher than those of both surfaces that were only plasma-treated without prior irradiation. This indicates a difference in work function at the surface, which leads to a global shift of the BE of the core levels. Indeed, when measuring the work function of a degreased sample prior and after plasma oxidation, we found an increment of 0.7\,eV. This has to be kept in mind when comparing relative differences in the core level binding energies for the samples.
The degreased samples exhibit the presence of hydroxide and hydrocarbon surface adsorbates with a state in the O\,1s spectrum at 531.2\,eV in addition to the metal-oxide peak at 530.1\,eV. The Cr\,2p$_{3/2)}$ state of the degreased samples exhibits a minor metallic component (Cr$^0$) around 574\,eV \cite{Kelly1998}, while the main chromium content is in the form of Cr$_2$O$_3$ (Cr$^{+3}$) with a representative peak at 576.7\,eV \cite{BIESINGER20112717,KADARI20173124}. For the irradiated sample, the carbon layer and other adsorbates were efficiently  removed by the plasma treatment and the surface is composed of Fe$_2$O$_3$ and Cr$_2$O$_3$. However, the plasma process even further oxidized the non-irradiated steel surface. While the Fe\,2p spectra are comparable, the dominating state at 578.6\,eV in the Cr\,2p$_{3/2}$ spectrum, additional to the shifted Cr$^{+3}$ feature of low intensity at 575.9\,eV, indicates the formation of CrO$_3$ (Cr$^{+6}$) at the surface \cite{RAHMAN1995203,ARONNIEMI2005108}.

When considering that the activated oxygen species of the plasma can react with carbon atoms at the surface or break Fe--C and Cr--C bonds to form volatile CO or CO$_2$, the effective removal of the carbon layer and the hydrocarbon molecules is assured. Simultaneously, Fe and Cr surface atoms with free valence electrons can react with oxygen atoms, leading to an increment of the oxidation state up to Fe$^{+3}$ and Cr$^{+6}$. On the bare stainless steel samples, this is well demonstrated by the measurements. However, for the irradiated sample, the oxygen plasma could not directly interact with the metal surface but was first etching the carbon layer during the plasma process. The process time seems to have been  sufficient to remove this layer completely, while the surface chromium atoms did not undergo full oxidation, as it is the case for the exposure of the non-irradiated samples to the plasma.

\section{Conclusions}
\label{sec:conclusions}
Carbon contamination has been identified upon low energy irradiation with H ions, which translates in major concerns for accelerating structures, as surface carbon has been associated, independently of the material under testing, as the main cause for reduction of the maximum achievable electric surface fields.
This study has demonstrated the efficiency of OPC in removing carbon contamination from metal surfaces. 
For the preliminary study case of CuBe2, it has been observed that the treatment was less effective, due to the fact that the electrodes were pre-exposed to vacuum arcing events while presenting carbon on the surface. However, for newly irradiated electrodes of Cu-OFE and SS316LN, the treatment successfully decreased the presence of carbon to levels acceptable for ultra high-vacuum applications. 
When tested for breakdown performance, treated CuBe2 and Cu-OFE showed electric field capabilities similar to surfaces that have not undergone irradiation beforehand, assuring that irradiation exposure and hydrogen blistering do not affect electric field performance. This is also true for SS316LN electrodes, where oxygen species from OPC have also caused a strong oxidation of the surface in the form of iron and chromium oxides. The strong oxidation in SS316LN samples after OPC seems to enhance breakdown resistance and enable the holding of higher electric surface fields, possibly through the increase of the work function. 

Further testing will be necessary to explore the role of such oxidation in stainless steel surfaces when studying breakdown events. To date, planning is being developed to assess SS316LN plain electrodes with OPC (without any irradiation exposure) in comparison with untreated ones, both as cathode and anode, for future DC testing, including detailed electron field emission measurements.  
Future work will also focus on optimizing OPC parameters to further enhance the efficiency of carbon removal.
At last, this research underlines the potential of oxygen plasma treatment as a key technology to be developed as a long-term \textit{in-situ} operational set-up inside particle accelerators, assuring the cleanliness and maintenance of inner surfaces of RF devices that could prolong their usable lifetime and even increase their performance.

\section{Acknowledgments}

The authors would like to express their gratitude to all LINAC4 personnel from the Beams Department at CERN, who have made these additional irradiation campaigns possible, with a special thanks to Edgar Sargsyan for his support.
The authors would also like to thank Bernard Henrist for his help in manipulating the plasma treatment set-up. 
A special acknowledgement is due to the Metrology and Non-Destructive Testing team at CERN, for the sharing of knowledge and for giving access to the microscopy laboratory. Additionally, we thank the members of the Mechanical Workshop Facility, responsible for the machining and metrology of the tested electrodes.

\section{Conflict of interest}
The authors declare that the research was conducted in the absence of any commercial or financial relationship that could be construed as a potential conflict of interest.

\bibliographystyle{unsrt}
\bibliography{bib_main}

\end{document}